\documentclass[aps,pra,twocolumn,showpacs,groupedaddress]{revtex4}  
\usepackage{graphicx}  
\usepackage{dcolumn}   
\usepackage{bm}        
\usepackage{amssymb}   
\usepackage{amsmath}
\usepackage{float}
\usepackage{ulem}
\usepackage{enumerate}
\begin{document}
\title{Spontaneous symmetry breaking due to the trade-off between \\ attractive and repulsive couplings}
\author{K. Sathiyadevi$^{1}$, S. Karthiga$^{2}$, V. K. Chandrasekar$^{1}$,  D. V. Senthilkumar$^3$ and M. Lakshmanan$^{2}$}
\address{$^1$Centre for Nonlinear Science \& Engineering, School of Electrical \& Electronics Engineering, SASTRA University, Thanjavur -613 401, Tamil Nadu, India. \\$^2$Centre for Nonlinear Dynamics, School of Physics, Bharathidasan University, Tiruchirappalli - 620 024, Tamil Nadu, India.\\ 
$^3$School of Physics, Indian Institute of Science Education and Research, Thiruvananthapuram-695016, India.}
\begin{abstract} 
\par Spontaneous symmetry breaking (SSB) is an important phenomenon observed in various fields including physics and  biology.   In this connection, we here show that the trade-off between attractive and repulsive couplings can induce spontaneous symmetry breaking in a homogeneous system of coupled oscillators.  With a simple model of a system of two coupled Stuart-Landau oscillators, we demonstrate how the tendency of attractive coupling in inducing in-phase synchronized (IPS) oscillations and the tendency of repulsive coupling in inducing out-of-phase synchronized (OPS) oscillations compete with each other and give rise to symmetry breaking oscillatory (SBO) states and interesting multistabilities.    Further, we provide explicit expressions for synchronized and anti-synchronized oscillatory states as well as the so called oscillation death (OD) state and study their stability.   If the Hopf bifurcation parameter (${\lambda}$) is greater than the natural frequency ($\omega$) of the system, the attractive coupling favours the emergence of an anti-symmetric OD state via a Hopf bifurcation whereas the repulsive coupling favours the emergence of a similar state through a saddle-node bifurcation.  We show that an increase in the repulsive coupling not only destabilizes the IPS state but also facilitates the re-entrance of the IPS state. 
\end{abstract}
\maketitle
\section{Introduction}
\par Complex patterns are observed in a wide variety of natural systems including physical, biological and chemical systems \cite{pik1,pik2,pik3,zou1,heart,kurths,saxena}.  A system of coupled oscillators serves as an excellent framework to unravel and to enhance our understanding on the underlying
dynamics of many complex systems.  For example, studies revealed that the large scale synchronization observed in neural networks are linked to several neurological diseases like essential tremor and tremor in Parkinson's disease \cite{pik1,pik2,pik3}.  Similarly, the suppression of normal sinus rhythm of pacemaker cells and other oscillation suppressions can now be understood in terms of the interaction of oscillators in the network \cite{zou1,heart}. Oscillation death (OD) observed in coupled oscillators has been interpreted as a background mechanism of cellular differentiation and amplitude death is being used as a mechanism for stabilization of physical or chemical systems \cite{kurths,saxena}.
\par  Spontaneous symmetry breaking (SSB) \cite{stroc} is a phenomenon  that can facilitate the onset of a rich variety of complex patterns observed  in several natural systems.  In SSB, asymmetric states arise from symmetric systems spontaneously as a control parameter is varied.  In other words,  the resultant asymmetric states do not show invariance under certain symmetry operations despite the equations of motion of the system exhibiting such an invariance.  SSBs can be widely observed in various natural systems including physical, biological and chemical systems \cite{stroc, pra_sk, pnas,pre,essay,cel1,symme,turing, sita}.  In the physical context, the understanding of SSB is central to the development of particle physics and many body theory \cite{stroc}.  Considering biological systems, SSB is crucial for cell movement, polarity and developmental patterning and is closely related to functional diversification on every scale, from molecular assemblies to subcellular structures, cell types themselves, tissue  architecture, and embryonic body axes \cite{essay}.  The phenomenon of SSB  helps in the formation of Turing patterns in organisms \cite{turing}.  SSB also leads to the complex pattern formation in brain dynamics~\cite{sita}. 
\par Among the various types of interactions considered in the literature, attractive (excitatory) and/or repulsive (inhibitory) couplings  are found in a variety of biological, chemical and physical systems. For example, in the case of neural networks \cite{pnas_1,nagu}, the suprachiasmatic nucleus in the brain is proposed to have attractive and repulsive couplings \cite{pnas_1} and in neurons \cite{nagu} excitatory and inhibitory synaptic couplings are known to exist. The combination of positive and negative feedbacks can be seen in genetic networks \cite{r6,gene_fur} as well.     
\par In this manuscript, we consider a system of two coupled identical oscillators, namely the paradigmatic Stuart-Landau limit cycle oscillators, with both  attractive and repulsive couplings, and investigate the effect of the trade-off between them resulting in a rich variety of dynamical behaviors and interesting multistabilities.  The attractive coupling is known to have the tendency to align the oscillators in an in-phase synchronized state (IPS) \cite{book_pik}.  In contrast, the repulsive coupling has the tendency to align the oscillators in an out-of-phase synchronized state (OPS) \cite{book_pik}.  We here deduce the explicit expressions of these states, namely IPS, OPS and also OD states and study their stability with respect to the attractive and repulsive coupling strengths.  It is to be noted that the explicit expressions for the IPS and OD states of coupled Stuart-Landau oscillators are well reported \cite{sec_od, prema, zakr} whereas the explicit expression for the OPS state has not yet been reported for coupled dynamical systems other than the phase only models.  Further, with numerical analysis, we show the existence of SSB state due to the trade-off between the attractive and repulsive couplings in a homogeneously coupled system.  Also, we demonstrate that the attractive coupling favours the emergence of an anti-symmetric OD state via a Hopf bifurcation whereas the repulsive coupling favours the emergence of a similar state through a saddle-node bifurcation.    We also find the re-entrance of in-phase synchronized state as the strength of the repulsive coupling is increased, which is a counter-intuitive behavior.    
\par The plan of the paper is as follows.  In Sec. \ref{model}, we present the model under consideration.  In Sec. \ref{antica}, we will investigate the existence and stability of different states in the symmetric and anti-symmetric subspaces.   In Sec. \ref{spont}, we illustrate the bifurcations leading to oscillation death state and elucidate the appearance of symmetry breaking oscillations through the trade-off between attractive and repulsive couplings.   Then in Sec. \ref{odsup}, we consider the spontaneous symmetry breaking OD state in the attractively coupled system and show that the introduction of repulsive interaction destabilizes the spontaneous symmetry breaking OD state.  Finally in Sec. \ref{conc1}, we summarize  the above results.
\section{\label{model}The Model}
\par We consider the coupled version of a simple, paradigmatic model, namely the Stuart-Landau oscillator \cite{book_pik}, representing a normal form of the Hopf bifurcation \cite{kuram, norm}.  It is known that weakly nonlinear oscillators can be modeled by the Stuart-Landau equation \cite{kuram} near Hopf bifurcation.  For example, the usefulness of such a model in studying neural networks has been explored in \cite{lbio1, lbio2}.  A system of two coupled Stuart-Landau oscillators with combined attractive and repulsive couplings is represented by
\begin{eqnarray}
\dot{z_j}=f(z_j) +\epsilon_1 ({\mathrm{Re}}[z_k-z_j]) -i\epsilon_2({\mathrm{Im}}[z_k-z_j]),  
\label{mod}
\end{eqnarray}
where the state variables $z_j=x_j+i y_j \in C$, $f(z_j)=(\lambda+i \omega -|z_j|^2)z_j$, $j,k=1,2$ and $k \neq j$.  In (\ref{mod}),  $\lambda$ and $\omega$ correspond to the Hopf bifurcation parameter and natural frequency of the systems, respectively.  Note that the attractive coupling (positive feedback) among the identical oscillators is established through the variables $x_j$, while the repulsive coupling (negative feedback) is achieved through the variables $y_j$.  
\par The emerging dynamics of the system (\ref{mod})  in the presence of either the attractive coupling alone \cite{kurths, sec_od} or the repulsive coupling \cite{repul1, repul2, repul3} has been well studied.  Efforts have also been taken to study the underlying dynamics in the presence of both attractive and repulsive couplings mostly in the phase  oscillators~(cf. \cite{daido1, strog, strog2, iat, spin, s_ana, kur}), which include
the conformist-contrarian  models \cite{strog, strog2},  models with spin glass type interactions \cite{spin} and models with dynamically varying attractive and repulsive interactions or adaptive interaction \cite{s_ana, kur}.
In contrast, we consider both the amplitude and phase effects in demonstrating our results. Considering such  general oscillators, only a very few works have been reported in the presence of both the attractive and repulsive couplings \cite{attr2, attr3, attr1} under different contexts.  The phenomenon of spontaneous symmetry breaking leading to heterogeneous dynamical nature (asymmetric states) due to the trade-off between the two couplings has not yet been demonstrated in any of these works.   
\par  In most of the earlier works (cf. \cite{daido1,strog, strog2}), the coupling is designed in such a way that few of the oscillators in the network experience attractive coupling while the remaining experience repulsive coupling.  In our case, both the systems in Eq. (\ref{mod}) are coupled with both attractive and repulsive interactions through different variables.  The homogeneously coupled system in (\ref{mod}) with attractive-repulsive interactions exhibits (i) permutational/translational symmetry $z_1 \rightarrow z_2$ and (ii) permutational parity symmetry $z_1 \rightarrow -z_2$.  In the following, we show that in a certain range of parameters, the dynamics of the homogeneous system (\ref{mod}) becomes heterogeneous due to the SSB.   We also note here that the attractive-repulsive couplings in (\ref{mod}) explicitly break the rotational symmetry present in the isolated Stuart-Landau oscillators. 
\section{\label{antica} Dynamics and stability of different states}
\par To study the dynamics of the considered coupled system, we first rewrite Eq. (\ref{mod}) in terms of the symmetric ($z_+$) and anti-symmetric ($z_-$) variables
\begin{eqnarray}
z_+=\frac{(z_1+z_2)}{2}, \qquad z_-=\frac{(z_1-z_2)}{2}.
\end{eqnarray}
Eq. (\ref{mod}) in terms of these new variables is given by 
\begin{eqnarray}
\dot{z}_+&=&\frac{1}{2}\left(f(z_++z_-)+f(z_+-z_-)\right), \nonumber \\
\dot{z}_-&=&\frac{1}{2}\left(f(z_++z_-)-f(z_+-z_-)\right) \nonumber \\ && \hspace{3cm} - 2\epsilon_1 \mathrm{Re}(z_-)+i 2 \epsilon_2 \mathrm{Im}(z_-). \quad \quad 
\end{eqnarray} 
In the in-phase subspace, $Z_+=\{(z_+, z_-)|z_- \equiv 0\}$, and in the anti-symmetric subspace, $Z_-=\{(z_+, z_-)|z_+ \equiv 0\}$.  Thus in the symmetric and anti-symmetric subspaces, the dynamical equations can be reduced, respectively, to
\begin{eqnarray}
\dot{z}_+=f(z_+) , \; \; \dot{z}_-=0,
\label{symsub}
\end{eqnarray}
and
\begin{small}
\begin{eqnarray}
\dot{z}_-=f(z_-) -2 \epsilon_1 \mathrm{Re}(z_-)+2 i \epsilon_2 \mathrm{Im}(z_-), \;\; \dot{z}_+=0. 
\label{ansub}
\end{eqnarray}
\end{small}
Note that the dynamical equation in the symmetric subspace is similar to the independent Stuart-Landau oscillator so that the periodic oscillations in this subspace are found to be identical to the one observed in the isolated  Stuart-Landau oscillator.  But in the anti-symmetric subspace, the orbits differ from the one observed in the isolated Stuart-Landau oscillator.   In the following, we present the explicit expressions for the different oscillatory states, steady states and their stabilities. 

\begin{figure}
\includegraphics[width=0.8\linewidth]{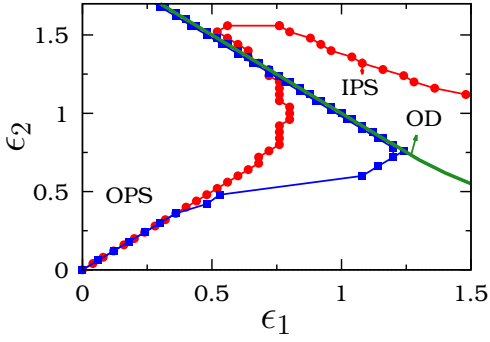}
\caption{(color online) Figure shows the boundaries of stable regions of IPS, OPS and OD states.  The area below the line (red line) with filled circles corresponds to the stable IPS state.  The area that lies between the $\epsilon_2$ axis and the curve (blue line) with filled squares represents the stable region of OPS state and the area that lies above the smooth line (green line) corresponds to the OD state.  The boundaries corresponding to IPS and OPS states are obtained through Floquet multipliers.  The boundary of the OD state is obtained from the linear stability of inhomogeneous steady states (\ref{anod}).  }
\label{figa}
\end{figure}
{\it (a) Dynamical states in symmetric subspace:}  Solving Eq. (\ref{symsub}) given above, the periodic orbits in the symmetric subspace can be written as
\begin{eqnarray}
z_+^*=\sqrt{\lambda} e^{i \omega t}.
\end{eqnarray}
Remembering $z_-=0$ in the symmetric subspace, we can write
\begin{eqnarray}
z_1^*=z_2^*=\sqrt{\lambda} e^{i \omega t}.
\label{sym1}
\end{eqnarray}
To know the stability of the above periodic orbit, we perturb it with slowly varying amplitudes in the form
\begin{eqnarray}
z_1=\sqrt{\lambda} e^{i \omega t}+\eta(t), \quad z_2=\sqrt{\lambda} e^{i \omega t}+\zeta(t),
\label{pert}
\end{eqnarray}
where $\eta(t)$ and $\zeta(t)$ are the perturbing terms and $\eta=\eta_1+i \eta_2 \in C$ and $\zeta=\zeta_1+i \zeta_2 \in C$.
Now substituting (\ref{pert}) in the system of equations (\ref{mod}) and by linearizing the resultant equations, we get
\begin{eqnarray}
\dot{\eta}&=(-\lambda+i \omega) \eta-\lambda \eta^* e^{2i \omega t}+\epsilon_1\mathrm{Re}(\zeta-\eta)-i\epsilon_2\mathrm{Im}(\zeta-\eta), \nonumber \\
\dot{\zeta}&=(-\lambda+i \omega) \zeta-\lambda \zeta^* e^{2i \omega t}+\epsilon_1\mathrm{Re}(\eta-\zeta)-i\epsilon_2\mathrm{Im}(\eta-\zeta). \quad 
\label{floeq}
\end{eqnarray}
\par { Integrating the above equation until $t=T=\frac{2 \pi}{\omega}$, we determine the Floquet multipliers $\rho_i$ ($i=1,\ldots,4$) from the fundamental matrix \cite{floq1,floq2}.  As long as, the four eigenvalues lie within the unit circle on the complex plane, the periodic orbit is stable.
 
\par From the Floquet multipliers that are obtained for different values of $\epsilon_1$ and $\epsilon_2$, we have depicted the boundary of stable regions of symmetric periodic orbits (in-phase synchronized state (IPS)) in Fig. \ref{figa}.  The area under the curve (line) with filled circles is the stable region of the IPS state.  From the figure, it is obvious that the introduction of $\epsilon_2$ shortens the stable region of IPS state and after a critical value of $\epsilon_2$, the IPS state is not stable for any value of $\epsilon_1$.}
 \par Other than the above symmetric periodic oscillations, a trivial steady state  ($z_1,z_2$)=($0,0$) is found to exist in the symmetric subspace,  which is unstable for all parametric values.  So it is not interesting physically. \\   \\
{\it  (b) Dynamical states in the anti-symmetric subspace:}   We have also deduced the solution of the corresponding dynamical equation in the anti-symmetric subspace (\ref{ansub}) with some effort as
\begin{eqnarray}
x_-(t)&=&\frac{e^{(\lambda_1+\bar{\epsilon})t} \cos(\theta)}{\left(C+e^{2t(\lambda_1+\bar{\epsilon})}(Q_0+Q_1 \cos (2 \theta)-Q_2 \sin(2 \theta))\right)^{\frac{1}{2}}}, \nonumber \\
y_-(t)&=&\frac{-1}{\omega}\left(\bar{\epsilon}-\sqrt{\omega^2-\bar{\epsilon}^2} \mathrm{tan}(\theta)\right) x_-(t),
\label{exact}
\end{eqnarray}
where $\theta=\sqrt{\omega^2-\bar{\epsilon}^2}\,t-\delta$, $\bar{\epsilon}=\epsilon_1+\epsilon_2$, $\lambda_1= \lambda-2\epsilon_1$ and $C$ and $\delta$ are integration constants.  The other constants $Q_0$, $Q_1$ and $Q_2$ are
\begin{align}
&Q_0=\frac{1}{\lambda_1+\bar{\epsilon}}, \quad Q_1=\frac{\bar{\epsilon} (\lambda_1 \bar{\epsilon} +\omega^2)}{\omega^2 (\lambda_1^2+2 \bar{\epsilon} \lambda_1+\omega^2)}, \nonumber \\ \quad &Q_2=\frac{\bar{\epsilon} \lambda_1 \sqrt{\omega^2-\bar{\epsilon}^2}}{\omega^2(\lambda_1^2+2 \bar{\epsilon} \lambda_1+\omega^2)}.
\label{Q_0Q_1Q_2}
\end{align}
 
 The solution in (\ref{exact}) is found to be periodic when $\omega > \bar{\epsilon}$.  In this case, we can write the state variables $x_i$ and $y_i$, $i=1,2,$ in the asymptotic limit ($t \rightarrow \infty$) as
\begin{small}
\begin{align}
&x_1^p(t)=\frac{\cos (\sqrt{\omega^2-\bar{\epsilon}^2}t)}{\left(Q_0+Q_1 \cos(2\sqrt{\omega^2-\bar{\epsilon}^2}t)-Q_2 \sin(2\sqrt{\omega^2-\bar{\epsilon}^2}t)\right)^\frac{1}{2}}, \nonumber \\
&y_1^p(t)=\frac{-1}{\omega}\frac{(\bar{\epsilon} \cos \theta-\sqrt{\omega^2-\bar{\epsilon}^2} \sin \theta)}{\left(Q_0+Q_1 \cos(2\sqrt{\omega^2-\bar{\epsilon}^2}t)-Q_2 \sin(2\sqrt{\omega^2-\bar{\epsilon}^2}t)\right)^\frac{1}{2}},
\label{anosc}
\end{align}
\end{small}
with $x_2^p=-x_1^p$ and $y_2^p=-y_1^p$.  
\par When $\omega<\bar{\epsilon}$, the solution in (\ref{exact}) implies that the system tends toward a steady state.  In this parametric range, it can be rewritten as
\begin{eqnarray}
x_-(t)&=&\frac{1}{\sqrt{2}}\frac{1+e^{-2 \theta'}}{D_1}, \nonumber \\
y_-(t)&=&\left[\frac{-\bar{\epsilon}}{\omega}+\frac{\sqrt{\bar{\epsilon}^2-\omega^2}}{\omega} \left( \frac{e^{-2\theta'}-1}{e^{-2 \theta'}+1}\right)\right] x_-(t),
\end{eqnarray}
where 
$D_1=\big[2C e^{-2t(\lambda_1+\bar{\epsilon})-2 \theta'}+2 Q_0 e^{-2 \theta'}+(Q_1-i Q_2)+(Q_1+i Q_2)e^{-4 \theta'}\big]^{\frac{1}{2}}$ and $\theta'=\sqrt{\bar{\epsilon}^2-\omega^2}t+i \delta$.  In the asymptotic limit $t \rightarrow \infty$, $x_-$ and $y_-$ tend to constant values leading to a pair of steady states given by
\begin{align}
x_1^*&=\pm \omega\left(\frac{(\lambda_1+\bar{\epsilon})+\sqrt{\bar{\epsilon}^2-\omega^2}}{2 \bar{\epsilon}(\bar{\epsilon}+\sqrt{\bar{\epsilon}^2-\omega^2})}\right)^{\frac{1}{2}}, \; \nonumber \\
y_1^*&=-\frac{(\bar{\epsilon}+\sqrt{\bar{\epsilon}^2-\omega^2})}{\omega} x_1^*,
\label{anod}
\end{align}
with $x_2^*=-x_1^*$ and $y_2^*=-y_1^*$.  In the above, the $\pm$ in $x_1^*$ appears due to the fact that if ($x_-(t), y_-(t)$) is solution of Eq. (\ref{ansub}), ($-x_-(t), -y_-(t)$) will also be the solution.  Stabilization of such inhomogeneous steady states  leads to the phenomenon of oscillation death (OD) \cite{saxena, zakr, revi2, cd1}.  
\par  Now we have to look at the stability of the above obtained states.  For this purpose, we perturb the anti-symmetric periodic solution as 
\begin{eqnarray}
x_1=x_1^p+\eta_1, \;\; y_1=y_1^p+\eta_2, \;\; \nonumber \\ x_2=-x_1^p+\zeta_1 \;\; y_2=-y_1^p+\zeta_2,
\end{eqnarray} 
where $\eta_i, \zeta_i$, $i=1,2$ are the perturbation terms. By substituting them in the system of equations (\ref{mod}) and by linearizing, we obtain
\begin{eqnarray}
\dot{\eta_1}=(\lambda -3 {x_1^p}^2-{y_1^p}^2)\eta_1-(\omega +2 x_1^p y_1^p) \eta_2+\epsilon_1(\zeta_1 -\eta_1), \nonumber \\
\dot{\eta_2}=(\lambda - {x_1^p}^2-3 {y_1^p}^2)\eta_2+(\omega -2 x_1^p y_1^p) \eta_1-\epsilon_2(\zeta_2 -\eta_2), \nonumber \\
\dot{\zeta_1}=(\lambda -3 {x_1^p}^2-{y_1^p}^2)\zeta_1-(\omega +2 x_1^p y_1^p) \zeta_2+\epsilon_1(\eta_1 -\zeta_1), \nonumber \\
\dot{\zeta_2}=(\lambda - {x_1^p}^2-3 {y_1^p}^2)\zeta_2+(\omega -2 x_1^p y_1^p) \zeta_1-\epsilon_2(\eta_2 -\zeta_2). 
\quad \label{floq22}
\end{eqnarray}
Note that the solution given in (\ref{anosc}) is periodic with respect to the period $\frac{2 \pi}{\sqrt{\omega^2-\bar{\epsilon}^2}}$.  The corresponding out-of-phase oscillations (OPS) are found to be stable in the area enclosed by the line with filled squares and the $\epsilon_2$ axis (see Fig. \ref{figa}). 
 
\par Whenever $\omega< \bar{\epsilon}$, the solution (\ref{exact}) tends to a pair of anti-symmetric steady states as given in (\ref{anod}).  We have studied the stability of these states and found that the corresponding eigenvalues are given by
\begin{eqnarray}
\mu_{1,2}&=&-2 {r^*}^2+\lambda\pm \sqrt{{r^*}^4-\omega^2} , \nonumber \\
\mu_3&=&-2 {r^*}^2, \quad \mu_4=-2 {r^*}^2-2\Delta \epsilon+ 2, \lambda
\label{odeig}
\end{eqnarray}
where ${r^*}^2={x_1^*}^2+{y_1^*}^2=\lambda-\Delta \epsilon+\sqrt{\bar{\epsilon}^2-\omega^2}$ and $\Delta \epsilon= \epsilon_1-\epsilon_2$.  
The stable region of such inhomogeneous steady states is also depicted in Fig. \ref{figa}. These steady states exist when $\epsilon_1>\omega-\epsilon_2$ (as the solution given in (\ref{exact}) does not represent oscillatory dynamics but represents a stable steady state).  From Fig. \ref{figa},  it is also clear that upon varying the value of $\epsilon_1$ there exists a direct transition from a stable anti-symmetric oscillatory state (OPS) to a stable OD state (indicated by a solid line) beyond a critical value of $\epsilon_2 (\approx 0.76)$.  On the contrary, for lower values of $\epsilon_2$ ($\epsilon_2<0.76$), these inhomogeneous steady states are not stabilized immediately upon destabilization of the OPS state.

\begin{figure}[htb!]
\begin{center}
\includegraphics[width=0.9\linewidth]{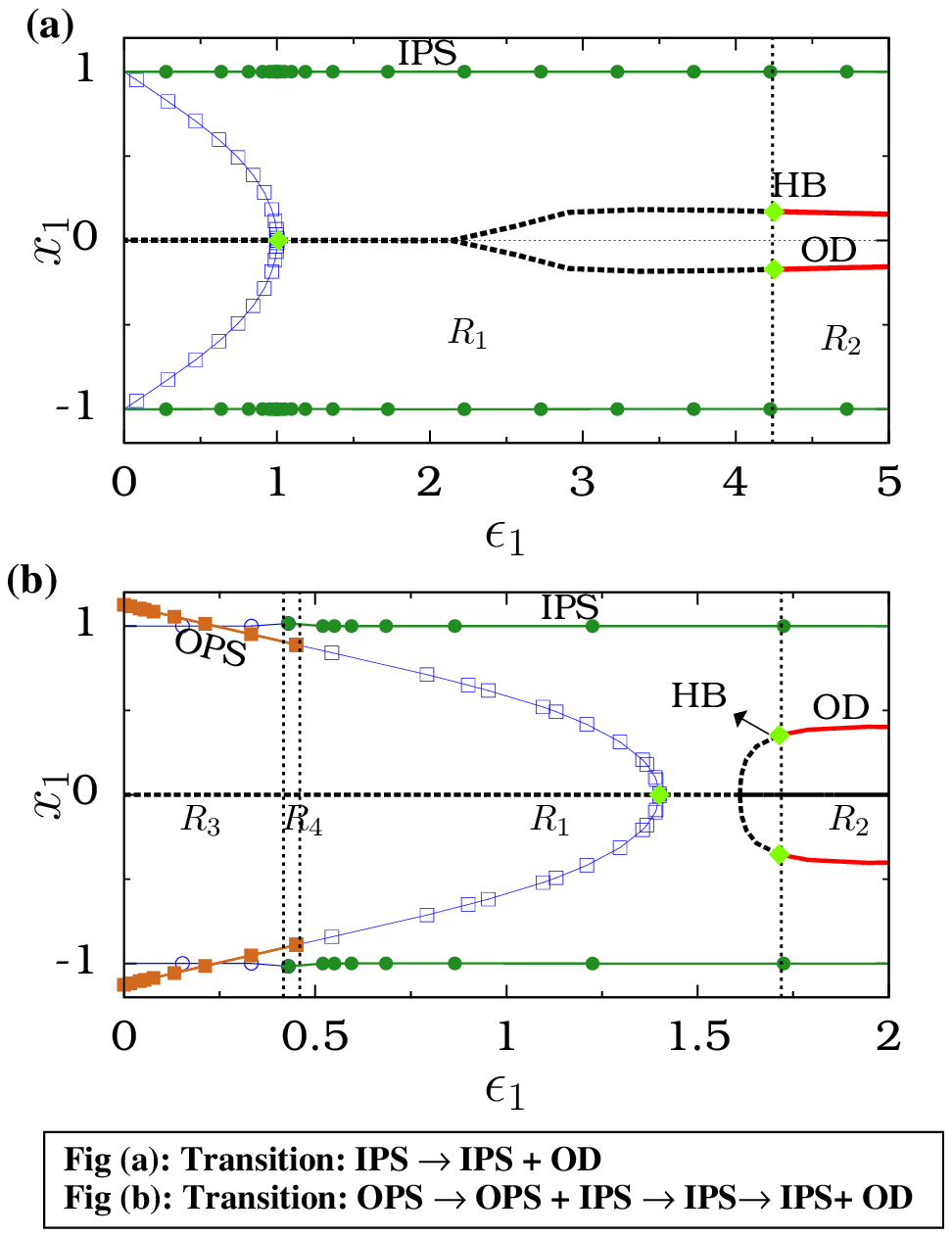}
\end{center}
\caption{(color online) Bifurcation diagrams of the system given in Eq. (\ref{mod}) for $\lambda=1.0$, $\omega=2.0$ and for (a) $\epsilon_2=0$, (b) $\epsilon_2=0.4$.  Here the bifurcation diagrams are obtained with respect to $\epsilon_1$, by using the software XPPAUT.  The lines with filled circle and square represent the maxima and minima of the stable IPS and OPS states, respectively, and the lines (blue colored) with empty circle and square represent the unstable nature of the IPS and OPS states, respectively.  The continuous (red) and dashed black lines, respectively, indicate the stable and unstable steady states.  HB represents the Hopf bifurcation point. $R_1$, $R_2$, $R_3$ and $R_4$ indicate different regions, where $R_1$ corresponds to stable IPS state, $R_2$ indicates multistability between IPS and OD, $R_3$ indicates stable region of OPS state and $R_4$ indicates the multistability between IPS and OPS.  }
\label{fig1}
\end{figure}
\section{\label{spont}Spontaneous symmetry breaking oscillations}

\par Theoretical studies in the earlier section deals only with explicit expressions for the states that exist in symmetric and anti-symmetric manifolds, whereas the explicit  expressions characterizing the existence of asymmetric states could not be deduced in the previous section.  However, while studying the dynamics of the  system  (\ref{mod}) numerically, we are able to observe that the system also has states that are asymmetric.  In connection with this, in this section we show the emergence of asymmetric states or spontaneous symmetry broken states  with suitable bifurcation diagrams.
\par  To begin with, we have depicted the bifurcation diagram of the system  (\ref{mod}) in the absence of the repulsive coupling ($\epsilon_2=0$) in Fig. \ref{fig1}(a), where the stabilization of the oscillatory branch in the range  $\epsilon_1$ $=$ ($1$, $5$) is shown.   This oscillatory branch refers to the IPS state given in  (\ref{sym1}) where its amplitude takes up the value $\sqrt{\lambda}$.

 Since the system loses its rotational symmetry while $\epsilon_1 \neq 0$ and $\epsilon_2=0$, increase in the value of $\epsilon_1$ in the region $R_2$ leads to a state which does not have rotational symmetry, namely the oscillation death (OD) state deduced in (\ref{anod}). Fig. \ref{fig1}(a) shows that this OD state stabilizes through a Hopf bifurcation. 

\begin{figure}[htb!]
\begin{center}
\includegraphics[width=0.95\linewidth]{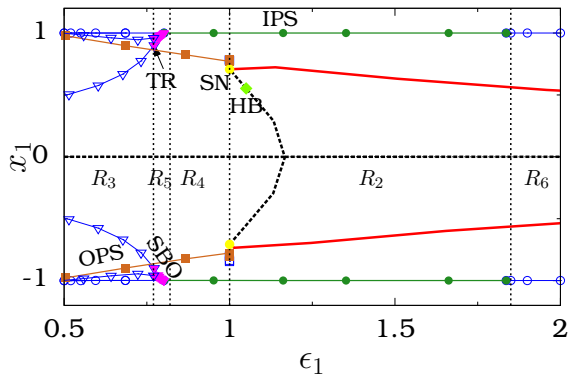}
\end{center}
\caption{(color online) Bifurcation diagram of the system given in Eq. (\ref{mod}) for $\lambda=1.0$, $\omega=2.0$ and for $\epsilon_2=1.0$. The stable and unstable IPS, OPS and OD states are represented as in Fig. \ref{fig1}.  Here, we see the appearance of a new branch, namely symmetry breaking oscillatory (SBO) branch which gets stabilized through an inverse torus bifurcation (TR).  The stable part of this branch is represented by lines (magenta colored) with filled triangles and their unstable nature is represented by empty triangles. Regions $R_2$, $R_3$ and $R_4$ have the same representation as given in Fig. \ref{fig1}. $R_5$ represents the stable region of SBO state and OD state alone is stable in $R_6$.}
\label{e2_1}
\end{figure} 
\begin{figure}[htb!]
\begin{center}
\includegraphics[width=1.0\linewidth]{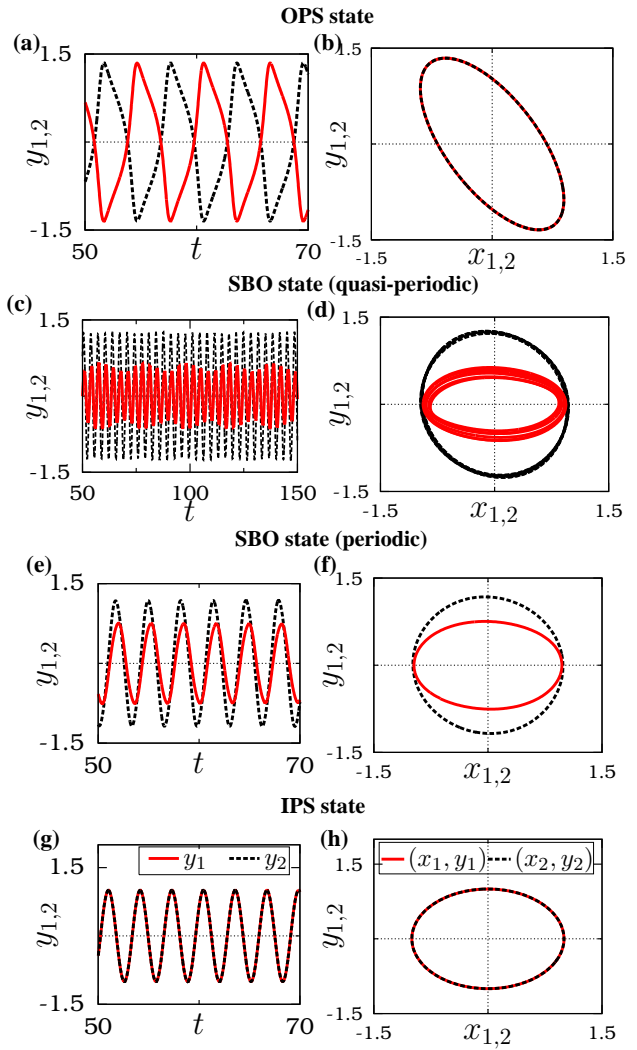}
\end{center}
\caption{(color online)  Figures (a)-(h) show the temporal behavior and phase portraits of the states observed in Fig. \ref{e2_1}, namely the OPS state, SBO state and IPS state.   These figures have been plotted for (a)-(b) $\epsilon_1=0.7$ (OPS state), (c)-(d) $\epsilon_1=0.77$ (SBO state - quasi-periodic oscillations), (e)-(f) $\epsilon_1=0.79$ (SBO state - periodic oscillations) and (g)-(h) $\epsilon_1=0.85$ (IPS state).}
\label{fig12}
\end{figure}

\begin{figure}[htb!]
\begin{center}
\includegraphics[width=0.95\linewidth]{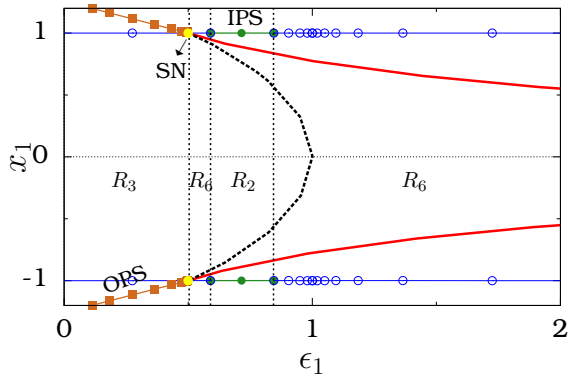}
\end{center}
\caption{(color online) Bifurcation diagram of the system given in Eq. (\ref{mod}) for $\lambda=1.0$, $\omega=2.0$ and for $\epsilon_2=1.5$. The stable and unstable nature of IPS, OPS and OD states are represented as in Figs. \ref{fig1} and \ref{e2_1}.  Similarly, the representations $R_2$, $R_3$ and $R_6$ are used as in Figs. \ref{fig1} and \ref{e2_1}. }
\label{e2_15}
\end{figure} 
\par Now by introducing a counteracting repulsive coupling, we have plotted the bifurcation diagram as a function of $\epsilon_1$ in Fig. \ref{fig1}(b) for $\epsilon_2=0.4$.  It shows the emergence of a new branch of stable oscillatory solution in the region $R_3$ and the temporal behavior of this oscillatory state confirms it to be an anti-phase or out-of-phase synchronized state (OPS) represented by Eq. (\ref{anosc}).  The repulsive coupling facilitates the emergence of OPS oscillations by destabilizing the IPS oscillations in the region $R_3$.  By increasing the strength of the attractive coupling, $\epsilon_1$, the stabilization of IPS oscillations can be seen in Fig. \ref{fig1}(b) and the emerging IPS oscillations are found to coexist with the OPS oscillations in the region $R_4$.  Further larger values of $\epsilon_1$ destabilizes OPS in the region $R_1$ and stabilizes the OD state in the region $R_2$ as evident from Fig. \ref{fig1}(b).

\par  Now we will discuss the observed dynamical transitions for further larger value of $\epsilon_2$, namely $\epsilon_2=1.0$.  We have plotted the bifurcation diagram illustrating the stable nature of various dynamical states in Fig. \ref{e2_1}. We can infer from Fig. \ref{e2_1} that the stable range of OPS state (indicated by lines connecting filled squares) is increased (it is found to be stable in the regions $R_3$, $R_5$ and $R_4$) and it touches the boundary of the OD region.  This elucidates that as noted during the theoretical analysis given in Sec. \ref{antica} that there exists a direct transition from anti-symmetric oscillatory state to anti-symmetric OD state where the OD state appears through a saddle-node bifurcation.  Thus, in contrast to the dynamical transition discussed in Fig. \ref{fig1}, here the strong repulsive coupling favours the onset of the OD state through a saddle-node bifurcation.  This is in contrast to the OD state which appears through a Hopf bifurcation for lower values of the repulsive coupling as shown in Fig. \ref{fig1}.  It is also evident from Fig. \ref{e2_1} that the range of the IPS state gets reduced and is found to be stable only in the regions $R_4$ and $R_2$ as a result of the repulsive coupling.  The stable region of the IPS state is suppressed not only for smaller values of $\epsilon_1$ but also for higher values of $\epsilon_1$ (Note that the branch corresponding to the IPS state is unstable not only in the regions $R_3$ and $R_5$ but also in the region $R_6$). Such a bi-directional destabilization of the IPS state is surprising,  as one would expect that the stability of this state in the lower range of $\epsilon_1$ alone will be affected by the increase in $\epsilon_2$.  But we observe a counter intuitive phenomenon in Fig. \ref{e2_1} where the IPS state is unstable  for larger values of $\epsilon_1$ also, that is in the region $R_6$.  This type of  destabilization of the IPS state destroys the multistability between the oscillatory IPS state and the OD state.
\par Another important dynamical behavior that can be observed from Fig. \ref{e2_1} is the one that arises before the stabilization of the IPS state.  In this region $R_5$,  there arises a new oscillatory branch (represented by magenta colored line with filled triangles in Fig. \ref{e2_1}) that has not been identified in our theoretical studies.  We label it as the symmetry breaking oscillatory (SBO) branch or asymmetric branch and is stabilized through an inverse torus bifurcation.  Due to the above bifurcation, quasi-periodic oscillations are found to co-exist with unstable SBO limit cycles near the boundary of $R_3$ with $R_5$ and at the bifurcation point TR, a transition from quasi-periodic oscillations to stable limit cycle oscillations occurs.    In the stable regions of quasi-periodic and periodic SBO oscillations, the permutational/translational symmetry ($z_1 \rightarrow \pm z_2$) of the system is broken spontaneously as will be elucidated in the following.  
\begin{figure*}[htb!]
\hspace{-0.2cm}
\includegraphics[width=1.0\linewidth]{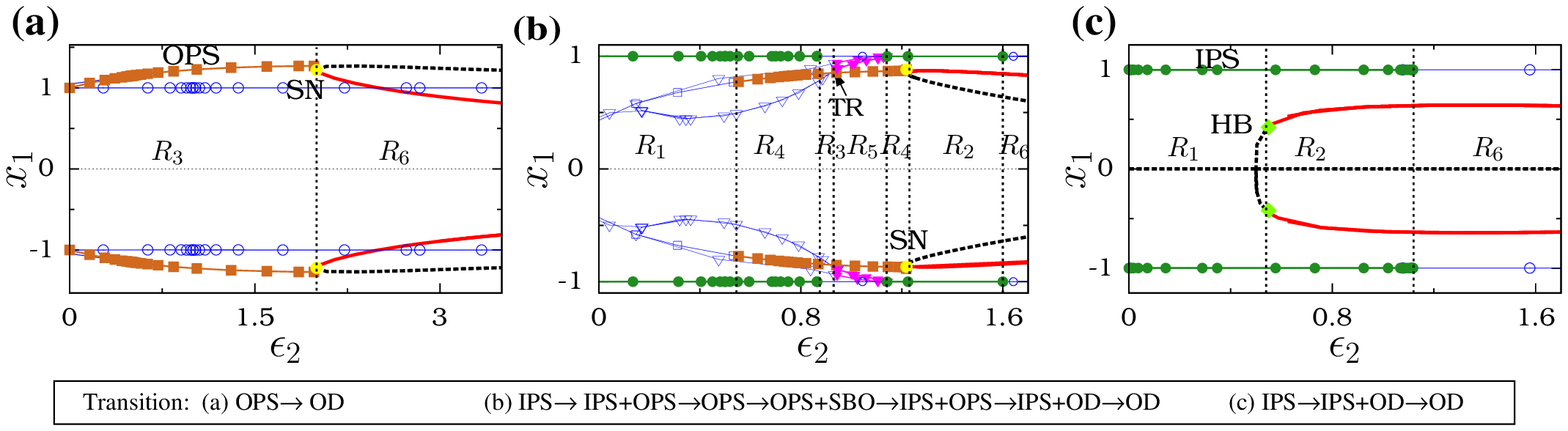}
\caption{(color online) Bifurcation diagrams for different values of $\epsilon_1$: (a) $\epsilon_1=0.0$, (b) $\epsilon_1=0.78$, (c) $\epsilon_1=1.5$ with $\lambda=1.0$ and $\omega=2.0$. The representation of different states such as IPS, OPS, SBO and OD are made similar to Figures \ref{fig1} and \ref{e2_1}.  The various regions are also named in a similar fashion used in the earlier figures.}
\label{fig2}
\end{figure*}
\par To demonstrate that the newly observed branch is attributed to symmetry breaking oscillations,  in Figs. \ref{fig12}(a)-(h) we have plotted the temporal behavior and phase portraits of all the oscillatory states observed in Fig. \ref{e2_1}.   The plots are arranged in the order in which they appear while increasing the value of $\epsilon_1$ in Fig. \ref{e2_1}.  We have depicted the temporal behavior and the phase portrait of the OPS state, respectively, in Figs. \ref{fig12}(a) and \ref{fig12}(b) for $\epsilon_1=0.7$.   From Fig. \ref{e2_1}, it is evident that the SBO state coexists with the OPS state in the range of $\epsilon_1$ $=$ ($0.77,0.81$).  
 Stabilization of SBO state occurs via the emergence of quasiperiodic oscillations at the boundary of $R_3$ and $R_5$,  as shown in Figs. \ref{fig12}(c) and \ref{fig12}(d). Inside the region $R_5$, this SBO state becomes periodic and its temporal behavior and the corresponding phase portrait are shown in Figs. \ref{fig12}(e)-\ref{fig12}(f) for $\epsilon_1=0.79$.  Further, increase in $\epsilon_1$ leads to the IPS oscillations as shown in Figs. \ref{fig12}(g)-\ref{fig12}(h) for $\epsilon_1=0.85$.

 \par  Now by comparing the temporal behaviors of different oscillatory states in Figs. \ref{fig12}(a)-\ref{fig12}(h), it is clear that the OPS oscillations and IPS oscillations preserve the $z_1 \rightarrow \pm z_2$ (or $x_1 \rightarrow \pm x_2$ and $y_1 \rightarrow \pm y_2$) symmetries of the system.  The exact matching of $x_1$-$y_1$ trajectory with the $x_2-y_2$ trajectory in Figs. \ref{fig12}(b) and \ref{fig12}(h) also corroborates the same.   On the other hand, for the SBO states,  Figs. \ref{fig12}(c) and \ref{fig12}(e) indicate that the amplitudes of $y_1$ and $y_2$ (also $x_1$ and $x_2$) are different from each other thereby elucidating the violation of permutational and permutational parity symmetries $z_1 \rightarrow \pm z_2$.  Further, Figs. \ref{fig12}(d) and \ref{fig12}(f) show that the trajectory $x_1-y_1$ not at all matches with that of $x_2-y_2$.  This type of heterogeneous dynamics in the homogeneously coupled system represents the underlying spontaneous symmetry breaking of the system.  As this state emerges by breaking the symmetry of the considered system given in Eq. (\ref{mod}), this state is called the symmetry breaking oscillatory (SBO) state. 
 
\par Increasing the value of $\epsilon_2$ to $1.5$, the associated bifurcation diagram is depicted in Fig. \ref{e2_15}.  Here the OPS oscillations that appear for lower values of $\epsilon_1$ lose their stability through saddle-node bifurcation and give rise to the OD state.  Further, the range of stable OPS state no longer widens for even larger $\epsilon_2$.  On the other hand, suppression of the OPS state is complemented with the spread of the stable OD region.  This is because on increasing the values of $\epsilon_1$ and $\epsilon_2$, the tendency of explicit rotational symmetry breaking dominates all the other observed dynamical states.  The stable range of IPS state is also suppressed to a large extent and it does not touch the boundary of the OPS oscillations in $R_3$ as can be seen in Fig. \ref{e2_15}.  As the SBO states arise at the boundary of the IPS oscillations with OPS oscillations (see $R_5$ in Fig. \ref{e2_1}), the SBO oscillations no longer exist in this case.  It is also to be noted that the anti-symmetric OD state appears through a saddle-node bifurcation even for lower values of $\epsilon_1$ than that in Fig. \ref{e2_1} for $\epsilon_2=1.0$.  This elucidates the fact that the repulsive coupling facilitates the transition from OPS to OD state through a saddle-node bifurcation. 
\par We have also depicted the bifurcation diagrams with respect to $\epsilon_2$, for various values of $\epsilon_1$  in Figs. \ref{fig2}(a)-\ref{fig2}(c).  We find from Fig. \ref{fig2}(a) that when $\epsilon_1=0$, there are two stable states, namely (i) anti-phase oscillations and (ii) OD state.  The transition from the former to the latter occurs through a saddle-node bifurcation.  Increasing the value of $\epsilon_1$ to $\epsilon_1=0.78$, we have plotted the bifurcation diagram in Fig. \ref{fig2}(b).  It is evident from the figure that only the IPS state is stable for lower values of $\epsilon_2$ and then in $R_4$, the OPS state gets stabilized along with the IPS state.   Further increase in $\epsilon_2$  destabilizes the IPS branch as seen in the region $R_3$ of Fig. \ref{fig2}(b).   After the region $R_3$, we find that in the region $R_5$, the asymmetric state gets stabilized along with the OPS state.  Increasing $\epsilon_2$ further, the IPS state again becomes stable by the destabilization of the SBO state.  Thus it leads to the reentrance of the IPS state as a function of $\epsilon_2$ which is explicitly dealt in Sec. \ref{reen}.  For further larger values of $\epsilon_2$, Fig. \ref{fig2}(b) shows that the OD state is the only stable state.  The bifurcation diagram for $\epsilon_1=1.5$ is depicted in Fig. \ref{fig2}(c), which clearly shows that the only stable states for this value of $\epsilon_1$ are (i) IPS and (ii) OD states.   By increasing $\epsilon_2$, we observe a transition from the IPS oscillatory state to the steady state through a Hopf bifurcation, whereas in the previous cases (in Figs. \ref{fig2}(a) and \ref{fig2}(b)) we observed transition from OPS state to OD state through a saddle-node bifurcation for lower values of $\epsilon_1$. 
\subsection{\label{competi}Trade-off between attractive and repulsive couplings in ($\epsilon_1,\epsilon_2$) space}
\par
\begin{figure*}[htb!]
\hspace{-0.2cm}
\includegraphics[width=0.9\linewidth]{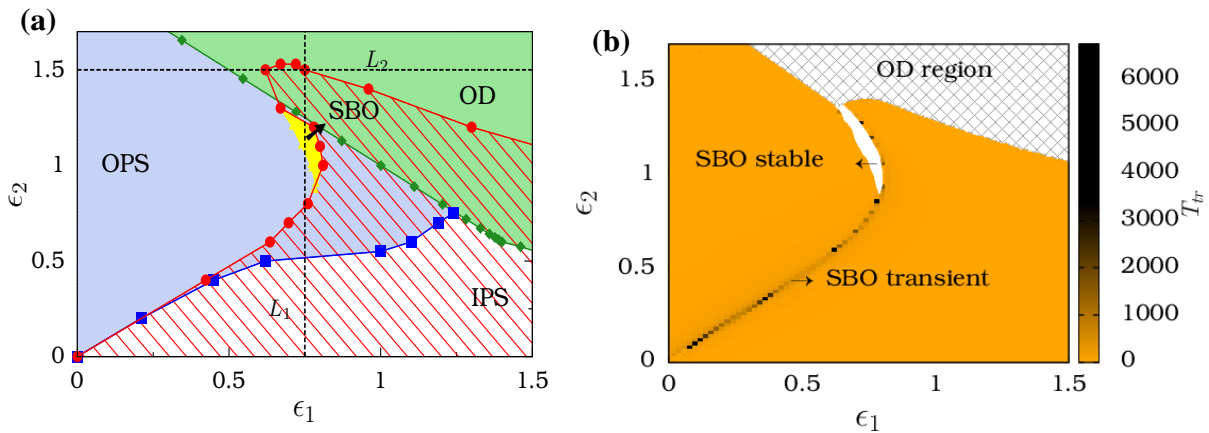}
\caption{(color online) Fig. (a): Stable regions of different states in the ($\epsilon_1$, $\epsilon_2$) parametric space.  In Fig. (b),  for different values of $\epsilon_1, \epsilon_2$, we have calculated the transient time ($T_{tr}$) that the system will take to reach either of the symmetric states, namely OPS and IPS oscillatory states by  noting the time at which the amplitudes of $z_1$ and $z_2$ become the same.  To obtain the Fig. (b), we have fixed the initial conditions as $z_1(0)=1.0+0.1 i$ and $z_2(0)=2.0+0.5i$. }
\label{fign3}
\end{figure*}

 From the numerical results, the stable regions of observed dynamical states are now illustrated in the ($\epsilon_1$, $\epsilon_2$) space in Fig. \ref{fign3}(a).  It clearly shows that the repulsive coupling does not favour the stabilization of IPS state and the attractive coupling does not favour the existence of OPS state.   But both the couplings favour the existence of anti-symmetric OD state, in the strong coupling limits.    
\par  Having known the tendencies of attractive and repulsive couplings, now we look into the competing effects of these two couplings in Fig. \ref{fign3}(a).  When both $\epsilon_1$ and $\epsilon_2$ ($\neq 0$) are small, the competition among the two opposing tendencies are weak so that an increase in $\epsilon_1$ for a particular lower value of $\epsilon_2$ causes the destabilization of OPS oscillations while it simultaneously stabilizes the IPS  oscillations.  On the contrary, an increase in $\epsilon_2$ for a particular value of $\epsilon_1$ gives rise to a destabilization of the IPS state and stabilization of the OPS state.  But when both $\epsilon_1$ and $\epsilon_2$ are increased, the competition among the attractive and repulsive couplings becomes strong.  Hence, while increasing $\epsilon_1$ for a particular large value of $\epsilon_2$, the OPS oscillations will not lose their stability at the onset of the IPS oscillations.  The OPS oscillations retain their stability after the IPS state becomes stable and so there arises  multistability among the OPS and IPS oscillations.  Thus the trade-off between attractive and repulsive couplings facilitates the coexistence of inherently contrasting oscillating states, namely the in-phase and out-of-phase oscillatory states.

 \par  On increasing the values of both $\epsilon_1$ and $\epsilon_2$ further, the competition among them becomes more intense resulting in an intricate dynamics.  In this case, in addition to the observed multistability between IPS and OPS states, we observe another interesting phenomenon, namely SSB.  In particular, in the range $\epsilon_2$ $\in$ ($0.89$, $1.3$), an increase in $\epsilon_1$ does not lead to the sudden appearance of IPS state in the OPS region giving rise to a multistability between the IPS and OPS states.  In this range of $\epsilon_2$, the permutational/translational symmetry of the coupled system (\ref{mod}) is broken spontaneously giving rise to symmetry broken oscillatory state before the IPS state gets stabilized.  
\begin{figure*}
\hspace{-1cm}
\includegraphics[width=0.8\linewidth]{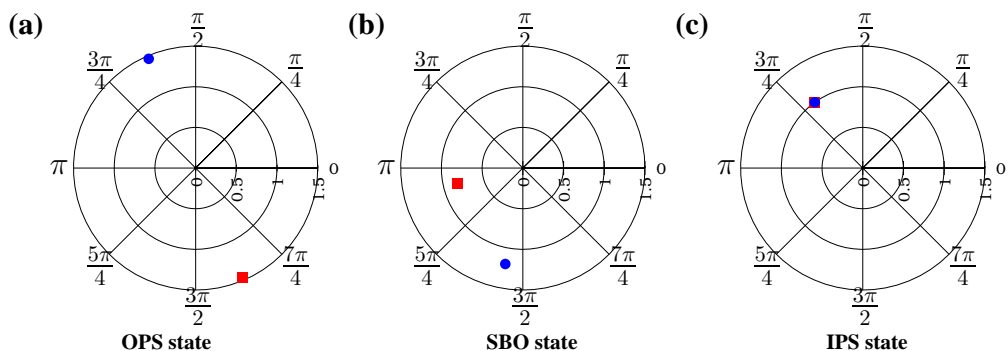}
\caption{(color online) The state of the two oscillators at a particular time in polar coordinates. Here, filled blue circles represent the state of the first oscillator and the filled red squares represent the state of the second oscillator at a particular time.  Fig. (a) is plotted for $\epsilon_1=0.7$, $\epsilon_2=1.0$, Fig. (b) is plotted for $\epsilon_1=0.77$, $\epsilon_2=1.0$ and Fig. (c) is plotted for $\epsilon_1=0.85$, $\epsilon_2=1.0$.  }
\label{phase}
\end{figure*}
\par  The above tendency of SSB not only exists for larger values of $\epsilon_1$ and $\epsilon_2$ but exists for lower values as well.  However, the SBO states are not stable in these regions.  Thus for lower values of $\epsilon_1$ and $\epsilon_2$, the asymmetric states are found to appear as transients along the boundary of the IPS oscillations with the OPS oscillations.   At this boundary, such asymmetric transient behavior persists for a considerably longer period of time.  To validate this observation, we have illustrated the transient time  (see  Fig. \ref{fign3}(b) ) taken by the system (\ref{mod})  to reach either the OPS or IPS oscillatory state starting from the fixed initial conditions $z_1(0)=1.0+0.1i$ and $z_2(0)=2.0+0.5i$.  Excluding the OD regions, it is evident from the figure that the transient time is lesser everywhere (yellow or shaded ones) except at the boundary of the IPS state with the OPS state which can be seen by a set of dark/black spots at their boundary in Fig. \ref{fign3}(b) corroborating the existence of larger transient region.  By comparing this curve with Fig. \ref{fign3}(a), it is evident that it lies at the boundary of the IPS state with the OPS state.  The unshaded areas in the Fig. \ref{fign3}(b) denote the stable regions of the SBO state where the system remains in this state over infinitely long time.  Thus it is clear from the above discussion that the trade-off between attractive and repulsive couplings leads to the manifestation of SSB in the coupled oscillators.

\par    In order to explain how the trade-off between the considered couplings result in the SBO states, we express the state variables $z_j$'s in polar form $z_j=r_j(t) e^{i \theta_j(t)}.$  In Figs. \ref{phase}(a)-\ref{phase}(c), we have depicted the snapshots of the system in terms of these polar coordinates for the OPS, SBO and IPS states, respectively.  It is known that the nature of the repulsive coupling is to separate the oscillators apart from each other.  In separating the two oscillators apart, the repulsive coupling finds a restriction implied by the symmetry of the underlying evolution equation (that is the permutational symmetries).  { Thus  the two oscillators are restricted to evolve in the same orbit but with $\pi$ phase difference.}  This can be seen clearly in Fig. \ref{phase}(a), a snapshot obtained for $\epsilon_2=1.0$ and $\epsilon_1=0.7$. This figure shows that the phase of the first (filled circle) and second (filled square) oscillators ($\theta_1$, $\theta_2$) are separated by an angle $\pi$ and the radius $r_1$ and $r_2$ are found to be the same.   In contrast, the attractive coupling tends to align the components of the coupled system to evolve in phase with each other.  Thus an increase in the value of the attractive coupling to $\epsilon_1=0.79$ tends to bring the two oscillators closer and it is evident from  Fig. \ref{phase}(b) that the phase difference among the oscillators is reduced. But the repulsive coupling strongly competes with the effect of the attractive coupling in this region and prevents the oscillators from evolving in-phase with each other.  Because of this trade-off between the repulsive and attractive couplings, the symmetry of the system is broken spontaneously for appropriate coupling strengths and renders $r_1$ and $r_2$ to be different in Fig. \ref{phase}(b).  Hence, we find the trajectories of $z_1$ and $z_2$ to be different in phase space as depicted in Figs. \ref{fig12}(d) and \ref{fig12}(f).  Increasing $\epsilon_1$ further, the attractive coupling becomes more dominant so that the two oscillators now follow the same path and their phases are also found to be the same as is evident from  Fig. \ref{phase}(c). 
\begin{figure}
\includegraphics[width=1.05\linewidth]{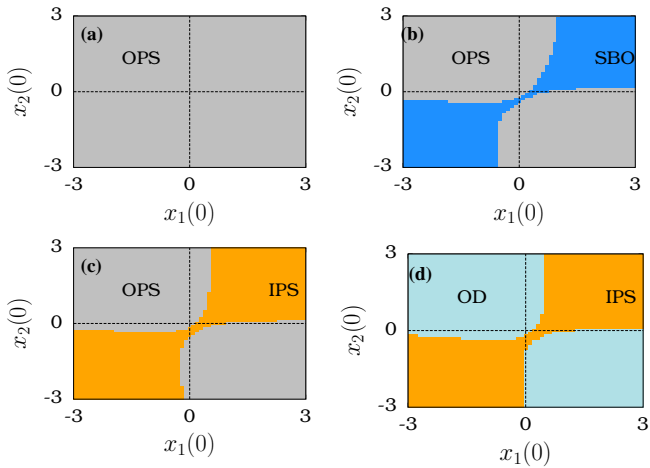}
\caption{(color online) The basins of attraction for different values of $\epsilon_1$ and for $\epsilon_2=1.0$, $\omega=2.0$ and $\lambda=1.0$:  Here, we have fixed $y_1(0)=0.5$ and $y_2(0)=0.2$ and found the basins of attraction of different states by varying $x_1(0)$ and $x_2(0)$. The Figs. (a)-(d) are plotted respectively for $\epsilon_1=0.75, 0.77,0.84$ and $1.0$. }
\label{bas}
\end{figure}

\par  It is also observed in  Fig. \ref{fign3}(a) that the OPS state is not destabilized with the stabilization of the SBO state.  In other words, the strong trade-off between the attractive and repulsive couplings leads to a symmetry broken state for only certain initial conditions and for other initial conditions the system tends towards the symmetric OPS state (Note that although the symmetric state is stable along with the asymmetric state, the symmetry in this parametric region is still said to be spontaneously broken).  Then there may arise a question that how can the OPS state retains its stability for certain initial conditions in the stable region of the SBO state and what are the initial conditions that lead to SBO and OPS states.  The answer to the question is as follows:  if the initial condition of the system is almost anti-symmetric (that is the regions in which the signs of $x_1(0)$ and $x_2(0)$ are opposite or that of $y_1(0)$ and $y_2(0)$ are opposite), the system can be easily stabilized to the OPS state where the tendency to align the oscillators to in-phase is weak.  But if the initial conditions are symmetric (the regions in which both $x_1(0)$ and $x_2(0)$ are of the same sign and $y_1(0)$ and $y_2(0)$ are also of the same sign), the tendency to align the oscillators to in-phase is strong so that for these initial conditions the trade-off leads to symmetry broken oscillations.  To illustrate these facts clearly, we have plotted the basins of attraction for different values of $\epsilon_1$ and for $\epsilon_2=1.0$ in Fig. \ref{bas}.   Here the OPS state is the only stable state for $\epsilon_1=0.75$ as all initial conditions lead to it as shown in Fig. \ref{bas}(a).  Now increasing $\epsilon_1$ to $0.77$, the SBO state  simultaneously becomes stable and here we find that the basins of attraction of the OPS state lies in the regions where $x_1(0)$ and $x_2(0)$ are anti-symmetric (that is, the second and fourth quadrants of $x_1(0)-x_2(0)$ space). But in the first and third quadrants, $x_1(0)$ and $x_2(0)$ are of the same sign (note that $y_1(0)$ ($=0.5$) and $y_2(0)$ $(=0.2)$ are also of the same sign) and so the tendency of aligning the oscillators in-phase is strong here, which leads to the SBO state.  Similarly, in the multi-stable region of IPS and OPS states, the basin of attraction of the IPS state lies in the same region where the basin of attraction of the SBO state exists.  To illustrate the above, we have increased $\epsilon_1$ to $0.84$ and $1.0$ and depicted the basins of attraction of the IPS and OPS states and that of the IPS and OD states  in Figs. \ref{bas}(c) and \ref{bas}(d), respectively.  From both the figures, we observe that the basin of attraction of the IPS state is concentrated in the first and third quadrants.  This shows that the tendency of aligning the oscillators to in-phase is strong in these regions of $x_1(0)-x_2(0)$ space.  This is the reason why we observe the stabilization of the SBO state for only such symmetric initial conditions and stabilization of the OPS state for other initial conditions.   
Thus it is clear from the above that the trade-off between the repulsive and attractive couplings breaks the symmetry of the system spontaneously and gives rise to SBO states. 

\begin{figure*}
\hspace{-1.0cm}
\includegraphics[width=0.78\linewidth]{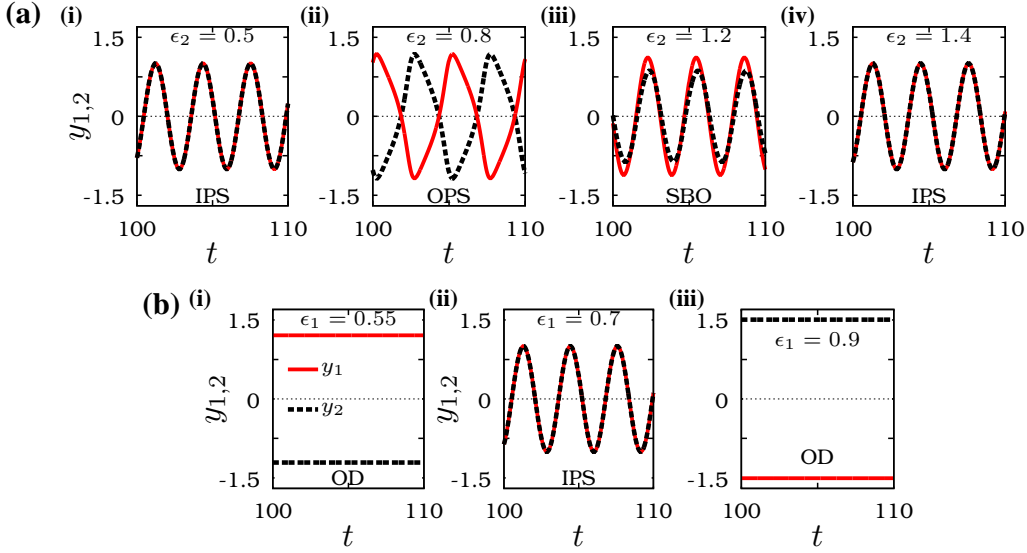}
\caption{(color online) Figs. (a)(i)-(iv) are plotted to demonstrate the reentrance of IPS state along the line $L_1$ in Fig. \ref{fign3}(a) for different values of $\epsilon_2$.  Figs. (b)(i)-(iii) are plotted by varying $\epsilon_1$ along the line $L_2$ in Fig. \ref{fign3}(a), showing the swing by mechanism exhibited by OD state. }
\label{temfig}
\end{figure*}
\begin{figure*}
\hspace{-1cm}
\begin{center}
\includegraphics[width=1.05\linewidth]{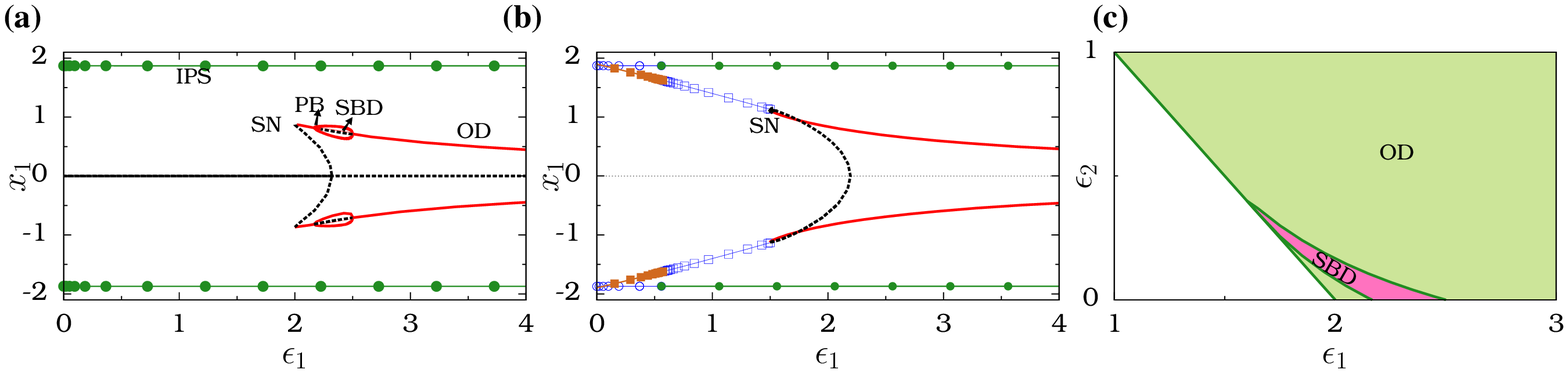}
\end{center}
\caption{(color online) Bifurcation diagrams for different values of $\epsilon_2$: (a) $\epsilon_2=0.0$, (b) $\epsilon_2=0.5$ for $\lambda=3.5$ and $\omega=2.0$. The OD state that appears after the pitchfork bifurcation in Fig. (a) are asymmetric and they are called the symmetry breaking death (SBD) state which is suppressed in Fig. (b). In the region of such asymmetric OD state, the anti-symmetric OD state is not stable. Due to the above, we have shown {the results of stability analysis on the stable regions of anti-symmetric OD state} in Fig. (c) which shows the stable region of SBD state (deep pink shaded region) inside the stable region of OD state (green shaded region).}
\label{fig3}
\end{figure*}
\subsection{\label{reen}Re-entrant synchronization}
\par  Another observation that can be inferred from Fig. \ref{fign3}(a) is the re-entrance of the in-phase synchronized state through the repulsive coupling.  This is evident from Fig. \ref{fign3}(a) when it is scanned along the line $L_1$.  We have also illustrated the above through the temporal behaviors of the system as a function of $\epsilon_2$ in Fig. \ref{temfig}(a).  The latter shows that the IPS oscillations which appear for $\epsilon_2=0.5$ (see Fig. \ref{temfig}(a)(i)) become destabilized by the increase of $\epsilon_2$, thereby leading to OPS state as it was shown in Fig. \ref{temfig}(a)(ii) for $\epsilon_2=0.8$.  Further increase in $\epsilon_2$ makes the IPS state (Fig. \ref{temfig}(a)(iv)) to reappear after the SBO state (Fig. \ref{temfig}(a)(iii)).    It is well known that the repulsive coupling has the tendency to oppose the IPS oscillations whereas here we observe that the increase in the repulsive coupling gives rise to the IPS state.  This re-entrance of the IPS state can also be observed from Fig. \ref{fig2}(b).  Such a re-entrance of a dynamical state as a function of a parameter is referred to as a swing by mechanism by Daido {et al} \cite{daido} in the presence of non-isochronicity. 

\par A similar swing like behavior can also be observed with reference to the OD state while we scan along the line $L_2$ in Fig. \ref{fign3}(a), where we find that for $\epsilon_1=0.55$, OD occurs as shown in Fig. \ref{temfig}(b)(i).  An increase in $\epsilon_1$ stabilizes the IPS state (in addition to the stable OD state) as shown in Fig. \ref{temfig}(b)(ii) for $\epsilon_1=0.7$.  Further increase in $\epsilon_1$ causes a destabilization of the IPS state and the OD alone is stable as shown in Fig. \ref{temfig}(b)(iii).  Thus for appropriate initial conditions near the basin of attraction of the IPS oscillatory state, a swing like behavior is observed in the OD state by varying $\epsilon_1$ as shown in Figs. \ref{temfig}(b).  

\section{\label{odsup}Suppression of asymmetric OD state for $\lambda>\omega$}
\par In the absence of repulsive coupling, it has been shown recently that there exists spontaneous symmetry breaking OD state for $\lambda>\omega$.  Such a state has been called the secondary OD state in \cite{sec_od}.  The appearance of such an asymmetric steady state has also been shown in Fig. \ref{fig3}(a) for $\omega=2.0$ and $\lambda=3.5$.  In this figure, the anti-symmetric OD state is found to appear through saddle-node bifurcation as a function of $\epsilon_1$.  Such an OD state soon loses its stability through a pitchfork bifurcation and it stabilizes the asymmetric fixed points which are of the form $(x_1^*, y_1^*, x_2^*, y_2^*)=(a_1^*, b_1^*, a_2^*, b_2^*)$, where $a_1^* \neq \pm a_2^*$ and $b_1^* \neq \pm b_2^*$.  These asymmetric fixed points also break the symmetry of the system spontaneously and these states can be called as symmetry breaking death (SBD) states.   We here study whether the introduction of $\epsilon_2$ can support this asymmetric state or not. For this purpose, we set $\epsilon_2=0.5$ and explore the bifurcation diagram in Fig. \ref{fig3}(b).  The figure clearly shows the disappearance of the SBD state and confirms that the introduced $\epsilon_2$ does not support the SBD states. We have also made sure of the above statement {through the theoretical results}, and in Fig. \ref{fig3}(c) we have plotted the stable range of the OD state {from the results of stability analysis} given in (\ref{odeig}).  The green shaded regions in Fig. \ref{fig3}(c) represent the stable OD region. Fig. \ref{fig3}(c) shows that inside the green shaded region, there exists a region (pink shaded) in which the OD state is not stable.  Now by comparing it with Fig. \ref{fig3}(a), we find that this pink shaded region corresponds to the stable region of the SBD state.  Thus the figure clearly proves the suppression of stable SBD region with an increase of $\epsilon_2$.
 
\section{\label{conc1} Summary}
\par In this article, we have considered a simple paradigmatic model of two coupled Stuart-Landau limit-cycle oscillators with attractive and repulsive couplings and investigated the dynamical  behaviors as a result of the competing effects of the two couplings.  The system of coupled Stuart-Landau oscillators studied in this paper has permutational symmetries and these symmetries were found to be spontaneously broken in a certain parametric region.  We have shown that the underlying reason for the appearance of such asymmetric states is the trade-off  between the attractive and repulsive couplings, where the tendency of inducing in-phase oscillations competes with the tendency of inducing out-of-phase oscillations.    We have also shown the appearance of  multi-stabilities between OPS, SBO and IPS oscillations.   Further, we have shown that for $\lambda<\omega$, the attractive coupling favours the emergence of anti-symmetric OD state  via a Hopf bifurcation whereas the repulsive coupling favours the emergence of anti-symmetric OD state through a saddle-node bifurcation. We  have also found the re-entrance of the IPS state as the strength of the repulsive coupling is increased, which is a counter intuitive behavior, despite the suppression of the IPS state for lower values of the repulsive coupling.  Importantly, the explicit expressions of the dynamical states such as IPS, OPS and OD states have also been obtained to study their stability. 
\section*{Acknowledgement}
KS and DVS are supported by a SERB-DST Fast Track scheme for young scientists under Grant No. ST/FTP/PS-119/2013.  The work of VKC forms part of a research project sponsored by INSA Young Scientist Project under Grant No. SP/YSP/96/2014.  SK thanks the Department of Science and Technology (DST), Government of India, for providing an INSPIRE Fellowship.  The work of ML is supported by a NASI Senior Scientist Platinum Jubilee fellowship program.

\end{document}